\documentclass[10pt, conference]{IEEEtran}

\IEEEoverridecommandlockouts

\ifCLASSOPTIONcompsoc
\else
\fi
\newif\ifdraft
\drafttrue

\usepackage[table]{xcolor}
\usepackage{balance}
\usepackage[hyphens]{url}

\usepackage{silence}

\usepackage{comment}

\specialcomment{rimuovere}{}{}
\specialcomment{rewrite}{\begingroup\color{red}}{\endgroup}
\specialcomment{tnsm}{\begingroup\color{black}}{\endgroup}
\specialcomment{commentolungo}{\begingroup\color{gray}}{\endgroup}
\specialcomment{draft}{\begingroup\color{blue}}{\endgroup}
\specialcomment{leggibile}{\begingroup\color{teal}}{\endgroup}
\specialcomment{warning}{\begingroup\color{red}}{\endgroup}
\newcommand{\attenzione}[1]{{\color{purple}XXX #1 XXX}\xspace}
\newcommand{\commento}[1]{{\color{gray}#1}}
\newcommand{\surplus}[1]{\textcolor{gray}{#1}}

\newcommand{\inlineTODO}[1]{\textcolor{orange}{\textbf{TODO: #1}}}

\ifdraft
    \usepackage[]{hyperref} 

    \usepackage{todonotes}
    
\else 
    \usepackage[draft]{hyperref} 
    \usepackage[disable]{todonotes}
    
    \specialcomment{new}{}{}
	\specialcomment{warning}{}{}
	
	\excludecomment{rewrite}
	\excludecomment{warning}
	\excludecomment{commentolungo}
	\excludecomment{rimuovere}
	\renewcommand{\attenzione}[1]{}
	\renewcommand{\commento}[1]{}
	\renewcommand{\surplus}[1]{}
	\renewcommand{\inlineTODO}[1]{}
\fi

\newcommand{\gr}{\rowcolor[HTML]{EFEFEF}}

\newcommand{\mcr}[1]{
\begin{tabular}[c]{@{}c@{}}#1\end{tabular}}

\newcommand{\ienum}[1]{\begin{enumerate*}[label=(\textit{\roman*})] #1 \end{enumerate*}}

\definecolor{tomato}{RGB}{255, 99, 71}
\definecolor{mediumseagreen}{RGB}{60,179,113}
\definecolor{goldenrod}{RGB}{218,165,32}

\newcommand{\nsc}{\ensuremath{\mathtt{F}}}
\newcommand{\ntx}{\ensuremath{\mathtt{N_{tx}}}}
\newcommand{\nrx}{\ensuremath{\mathtt{N_{rx}}}}

\newcommand{\Cbf}{\ensuremath{\mathbf{c}}}
\newcommand{\complexnum}{\ensuremath{\mathbb{C}}}
\newcommand{\sr}{\ensuremath{\mathtt{S_{x}}}}
\newcommand{\wsize}{\ensuremath{\mathtt{W}}}

\newcommand{\amplitude}{\ensuremath{\mathbf{A}}}
\newcommand{\phased}{\ensuremath{\bm{\Delta\phi}}}
\newcommand{\phase}{\ensuremath{\bm{\phi}}}
\newcommand{\amphd}{\ensuremath{\mathbf{A\&}\bm{\Delta\phi}}}

\newcommand{\wifi}{Wi-Fi\xspace}

\newcommand{\mrows}[2]{\multirow{#1}{*}{#2}}
\newcommand{\mcols}[2]{\multicolumn{#1}{c}{#2}}

\newcommand{\dopp}[0]{DS}

\newcommand{\phd}[0]{PHD}

\newcommand{\amp}[0]{AMP}
\newcommand{\stft}[0]{STFT}

\definecolor{emb}{HTML}{A47864}
\definecolor{onecnn}{HTML}{36827F}
\definecolor{bigru}{HTML}{FFB570}
\definecolor{dense}{HTML}{899F6A}
\definecolor{dropout}{HTML}{A793B9}
\definecolor{maxpooling}{HTML}{FFD966}
\definecolor{stub}{HTML}{29C979}
\definecolor{flatten}{HTML}{7EA6E0}
\definecolor{concat}{HTML}{E6E6E6}
\definecolor{input}{HTML}{FFBE98}

\newcommand{\sage}{\textsc{SAGE}\xspace}

\newcommand{\mall}{\mathcal{M}_{all}}
\newcommand{\mcross}{\mathcal{M}_{X}}
\newcommand{\cfr}{\mathbf{h}}

\newcommand{\bothclass}{\mathcal{T}_{s}}
\newcommand{\onlyclass}{\mathcal{T}_{u}}

\usepackage[framemethod=tikz]{mdframed}
\usepackage{xcolor}

\definecolor{violet}{RGB}{138, 43, 226}
\definecolor{lightviolet}{RGB}{240, 230, 255}

\newmdenv[
  innerlinewidth=.5pt,
  roundcorner=5pt,
  linecolor=violet,
  backgroundcolor=lightviolet,
  skipabove=1.5pt,
  skipbelow=1.5pt,
  innertopmargin=3pt,
  innerbottommargin=3pt,
  innerleftmargin=3pt,
  innerrightmargin=3pt,
]{takehomebox}

\usepackage{makecell}
\usepackage{amsmath}
\usepackage{amssymb}
\usepackage{amsfonts}
\usepackage{caption}
\usepackage{booktabs}
\usepackage{array}
\usepackage{wasysym}
\usepackage{lscape}
\usepackage{pdflscape}
\usepackage{afterpage}
\usepackage{bm}
\usepackage{rotating}
\usepackage{pdflscape}
\usepackage[inline]{enumitem}

\usepackage{mathtools}

\usepackage{cleveref}
\usepackage{dirtytalk}
\usepackage{amssymb}
\usepackage{amsthm}

\usepackage{graphicx}
\usepackage{epsfig}
\usepackage{epstopdf}
\usepackage[font=small]{caption}
\usepackage{booktabs}
\usepackage{subfig}
\usepackage[english]{babel}
\usepackage{comment}

\usepackage[T1]{fontenc} 
\usepackage[utf8]{inputenc}
\usepackage{setspace}
\usepackage[square,sort,comma,numbers]{natbib} 
\usepackage{multirow}
\usepackage{bm}
\usepackage{xspace} 
\usepackage[10pt]{moresize}
\usepackage{balance}
\usepackage{threeparttable}
\usepackage{fontawesome5}

\usepackage{eso-pic}

\newcommand{\IEEEcopyrightnotice}{%
\AddToShipoutPictureFG*{%
  \AtPageLowerLeft{%
    \hspace{0.5in}%
    \raisebox{0.25in}{%
      \parbox{\dimexpr\paperwidth-1in\relax}{%
        \centering\scriptsize
        \copyright~2026 IEEE. Personal use of this material is permitted.
        Permission from IEEE must be obtained for all other uses, in any current
        or future media, including reprinting/republishing this material for
        advertising or promotional purposes, creating new collective works,
        for resale or redistribution to servers or lists, or reuse of any
        copyrighted component of this work in other works.
      }%
    }%
  }%
}%
}

\usepackage[acronym]{glossaries}  
\setkeys{glslink}{hyper=false} 

\newacronym{aci}{ACI}{Adjacent Channel Interference}
\newacronym{adam}{ADAM}{Adaptive Momentum Estimation}
\newacronym{adc}{ADC}{Analog-to-Digital Converter}
\newacronym{agc}{AGC}{Automatic Gain Control}
\newacronym{ai}{AI}{Artificial Intelligence}
\newacronym{aoa}{AoA}{Angle-of-Arrival}
\newacronym{ap}{AP}{Access Point}
\newacronym{api}{API}{Application Programming Interface}
\newacronym{awgn}{AWGN}{Additive White Gaussian Noise}

\newacronym{ber}{BER}{Bit Error Probability}
\newacronym{bss}{BSS}{Basic Service Set}

\newacronym{cfo}{CFO}{Carrier-Frequency Offset}
\newacronym{cnn}{CNN}{Convolutional Neural Network}
\newacronym{cots}{COTS}{Commercial Off-The-Shelf}
\newacronym{csi}{CSI}{Channel State Information}
\newacronym{csis}{CSIS}{CSI-Sensing}
\newacronym{csma}{CSMA}{Carrier-Sense Multiple Access}
\newacronym{cv}{CV}{Computer Vision}

\newacronym{doi}{DOI}{Digital Object Identifier}
\newacronym{dac}{DAC}{Digital to Analog Converter}
\newacronym{dft}{DFT}{Discrete Fourier Transform}
\newacronym{dl}{DL}{Deep Learning}
\newacronym{dnn}{DNN}{Deep Neural Network}

\newacronym{eth}{ETH}{Extremely High Throughput}

\newacronym{fdma}{FDMA}{Frequency-Division Multiple Access}
\newacronym{fft}{FFT}{Fast Fourier Transform}

\newacronym{gps}{GPS}{Global Positioning System}

\newacronym{har}{HAR}{Human Activity Recognition}
\newacronym{hbr}{HBR}{Human Behavior Recognition}

\newacronym{isac}{ISAC}{Integrated Sensing and Communications}
\newacronym{idft}{IDFT}{Inverse Discrete Fourier Transform}
\newacronym{ifft}{IFFT}{Inverse Fast Fourier Transform}

\newacronym{lsp}{LSP}{Latent Space Parameter}
\newacronym{lts}{LTS}{Long Training Sequence}
\newacronym{los}{LOS}{Line-of-Sight}
\newacronym{nlos}{NLOS}{Non-line of Sight}

\newacronym{mac}{MAC}{Medium Access Control}
\newacronym{mcs}{MCS}{Modulation and Coding Scheme}
\newacronym{mpc}{MPC}{Model Predictive Control}
\newacronym{mimo}{MIMO}{Multiple-Input Multiple-Output}
\newacronym{mumimo}{MU-MIMO}{Multi User-MIMO}
\newacronym{ml}{ML}{Machine Learning}
\newacronym{mlp}{MLP}{Multi-Layer Perceptron}
\newacronym{mmwave}{mmWave}{millimeter Wave}

\newacronym{nn}{NN}{Neural Network}
\newacronym{nic}{NIC}{Network Interface Card}

\newacronym{ofdm}{OFDM}{Orthogonal Frequency Division Multiplexing}

\newacronym{pdr}{PDR}{Packet Delivery Rate}
\newacronym{per}{PER}{Packet Error Rate}
\newacronym{pps}{PPS}{Pulse Per Second}

\newacronym{relu}{ReLU}{Rectified Linear Unit}
\newacronym{rss}{RSS}{Received Signal Strength}
\newacronym{rssi}{RSSI}{Received Signal Strength Indicator}
\newacronym{rf}{RF}{Radio Frequency}
\newacronym{rfid}{RFID}{Radio Frequency Identification}

\newacronym{sdr}{SDR}{Software Defined Radio}
\newacronym{sfo}{SFO}{Sampling-Frequency Offset}
\newacronym{sinr}{SINR}{Signal to Interference plus Noise Ratio}
\newacronym{sir}{SIR}{Signal to Interference Ratio}
\newacronym{siso}{SISO}{Single-Input Single-Output}
\newacronym{simo}{SIMO}{Single-Input Multiple-Output}
\newacronym{snr}{SNR}{Signal to Noise Ratio}
\newacronym{ssid}{SSID}{Service Set Identifier}
\newacronym{stft}{STFT}{Short-Time Fourier Transform}

\newacronym{tdma}{TDMA}{Time-Division Multiple Access}
\newacronym{tpr}{TPR}{True Positive Rate}
\newacronym{tof}{ToF}{Time-of-Flight}

\newacronym{vae}{VAE}{variational autoencoder}

\newacronym{lcsi}{L-CSI}{Linux 802.11n CSI}
\newacronym{ncsi}{N-CSI}{Nexmon CSI}
\newacronym{acsi}{A-CSI}{Atheros CSI}

\newacronym{be}{BE}{Biometrics Estimation}
\newacronym{bg}{BG}{Body Gestures Recognition}
\newacronym{cd}{CD}{Copresence Detection} 
\newacronym{cl}{CL}{Crowd Localization}  
\newacronym{fd}{FD}{Fall Detection}   
\newacronym{hg}{HG}{Hand Gestures Recognition}   
\newacronym{hhi}{HHI}{Human-to-Human Interaction Recognition}   
\newacronym{hid}{HID}{Human Identification}
\newacronym{il}{IL}{Indoor Localization}
\newacronym{pc}{PC}{People Counting}
\newacronym{sp}{SP}{Sign Language Poses Recognition}
\newacronym{ul}{UL}{User Localization}
\newacronym{wve}{WVE}{Walking Velocity Estimation}

\newacronym{xai}{XAI}{eXplainable Artificial Intelligence}

\newglossarystyle{mystyle}{

  \renewcommand*{\glsgroupheading}[1]{}

}

\makeglossaries
\usepackage{breqn}
\usepackage{units}
\usepackage{color, colortbl}
\definecolor{royalblue}{rgb}{0.26, 0.41, 1}


\begin{document}

\title{XAI2CSI: Interpreting CSI with eXplainable AI\\for Human Activity Recognition
}

\author{
\IEEEauthorblockN{
    Idio Guarino\IEEEauthorrefmark{1},
    Alfredo Nascita\IEEEauthorrefmark{2},
    Domenico Ciuonzo\IEEEauthorrefmark{2},
    Damiano Carra\IEEEauthorrefmark{3},
    Antonio Pescap\'e\IEEEauthorrefmark{2}
    }
    \IEEEauthorblockA{
    \IEEEauthorrefmark{1}\textit{University of Bologna, Italy},
    \IEEEauthorrefmark{2}\textit{University of Napoli Federico II, Italy},
    \IEEEauthorrefmark{3}\textit{University of Verona, Italy},
    }
    \IEEEauthorblockA{idio.guarino@unibo.it, damiano.carra@univr.it, \{alfredo.nascita, domenico.ciuonzo, pescape\}@unina.it, }

}

\maketitle

\IEEEcopyrightnotice

\begin{abstract}
\wifi \gls{csi} has emerged as a key enabler for device-free \gls{har}, enabling low-cost, unobtrusive sensing using existing communication infrastructure.
However, \gls{dl} models trained on \gls{csi} data often 
struggle to generalize across users, environments, and device setups due to the context sensitivity of wireless propagation.
Despite this challenge, limited attention has been devoted to understanding model decisions and generalization failures.
This paper introduces XAI2CSI, a framework that leverages \gls{xai} to analyze \gls{dl}-based \gls{csi} sensing systems.
XAI2CSI employs \sage, a model-agnostic explainability method, to quantify temporal, spectral, and spatial \gls{csi} contributions to \gls{har} decisions under nominal and cross-context evaluations on IEEE 802.11ax data.
%
Our analysis reveals that the considered \gls{dl} model exhibits limited robustness to unseen conditions due to an over-reliance on context-specific \gls{csi} patterns, causing models to misinterpret the underlying signal dynamics when deployment conditions change.
The proposed methodology and findings provide a reference framework to explore alternative solutions and guide the development of robust, transparent \wifi sensing systems.


\end{abstract}


\glsresetall

\section{Introduction}
\label{sec:intro}


In recent years, \gls{dl} has been increasingly adopted to leverage wireless signals for environmental sensing~\cite{tan2022}.
%
%
A rich source of information is the \gls{csi} from \wifi signals, which has become a key enabler for device-free \gls{har}~\cite{nirmal2021}.
%
By exploiting subtle disturbances induced by human movement on radio-frequency propagation, \gls{csi} enables low-cost and non-intrusive sensing using existing commercial-off-the-shelf communication hardware, thus avoiding wearable-device limitations such as limited battery life, user discomfort, and complex deployment.
%
Despite its promise, the \emph{main obstacle} to widespread adoption is \emph{generalizability}:
\gls{dl} models trained on the \gls{csi} data may \emph{struggle} when deployed across different users, environments, or device setups due to wireless propagation's high sensitivity to context~\cite{guarino2026}.
Multipath effects, where signals reflect off walls, furniture, and people, create a unique \gls{csi} ``fingerprint'' for each specific setting. Consequently, even minor changes to the physical environment alter this fingerprint, degrading model performance and preventing the development of a universal, ``one-size-fits-all'' model that can be trained once and deployed anywhere without extensive retraining.
Although numerous studies have documented these performance degradations, they largely treat generalization as a performance problem, leaving open a key question: \emph{why do \gls{csi}-based \gls{har} models fail under unseen conditions}?

Answering this question is challenging because \gls{dl} models operate as \emph{black boxes}, making it \emph{unclear} which \gls{csi} components drive successful predictions and how their contribution changes across deployment contexts.
%
\gls{xai} provides tools to interpret complex models by identifying the features that most influence their decisions.
However, despite the acknowledged challenges in \gls{csi}-based sensing, the systematic application of \gls{xai} to understand model behavior and generalization remains \emph{largely unexplored}.

This paper bridges this gap through \textbf{XAI2CSI}, a framework that applies \gls{xai} to analyze \gls{csi}-based \gls{har} under nominal and cross-context conditions.
%
Specifically, XAI2CSI leverages \sage~\cite{covert2020sage}, a novel model-agnostic explainability method, to quantify temporal, spectral, and spatial \gls{csi} contributions to \gls{har} decisions.
Through rigorous evaluations on IEEE 802.11ax data~\cite{cominelli2023}, we investigate the models' failures in cross-user scenarios by analyzing feature importance across the spatial, temporal, and spectral dimensions of the \gls{csi} input.
%
Our analysis reveals limited robustness to unseen conditions due to an over-reliance on context-specific \gls{csi} patterns, causing models to misinterpret the underlying signal dynamics when deployment conditions change.
%
%
These findings provide actionable insights toward robust, transparent, and interpretable \wifi sensing systems.
%
Our \textbf{contributions} are: \ienum{
\item we evaluate \gls{dl}-based \gls{har} models under standalone and combined \gls{csi} representations (amplitude, phase difference, and their combination) using 802.11ax hardware;
\item we perform systematic cross-domain evaluations across users, environments, and receiver node positions to assess model generalization;
\item 
we apply \sage to provide a comprehensive \gls{xai} analysis, characterizing temporal, spectral, and spatial \gls{csi}  contributions under distribution shifts.

}


The paper is organized as follows: Sec.~\ref{sec:rw} reviews related works. Sec.~\ref{sec:methodology} details our \gls{xai} methodology. Secs.~\ref{sec:setup} and \ref{sec:evaluation} present the setup and results, while Sec.~\ref{sec:conclusions} concludes it.

\renewcommand{\arraystretch}{1.2} 

\begin{table}[!h]
\centering
\scriptsize
\caption{ 
Related \gls{csi}-based \gls{har} works and positioning of this work.
}
\label{tab:rw}
\resizebox{.96\columnwidth}{!}{ 
\begin{threeparttable}

\begin{tabular}{@{}r c  c c c c c@{}}
\toprule

\mrows{2}{\textbf{Paper}} & \mrows{2}{\textbf{802.11}} & \mrows{2}{\textbf{CSI data}} & \mcols{3}{\textbf{Cross Eval.}} & \mrows{2}{\textbf{XAI}} \\

\cmidrule(lr){4-6}
& & & \faHome & \faUser & \faBroadcastTower &\\
\midrule
\gr \citet{yousefi2017} & n & \amp & \Circle & \Circle & \Circle &\Circle\\
\citet{zou2018}  & n & \amp & \Circle & \Circle & \Circle &\Circle\\
\gr \citet{wang2019} & n & \stft & \CIRCLE & \CIRCLE & \CIRCLE &\Circle\\
\citet{yang2019} & n &  \phd & \CIRCLE & \CIRCLE & \Circle &\Circle\\
\gr \citet{chen2019} & n & \amp & \CIRCLE & \Circle & \Circle &\Circle\\
\citet{yang2022efficient} & n & \amp & \Circle & \Circle  & \Circle &\Circle\\
\gr \citet{zhang2022fewshot} & n & \amp & \CIRCLE & \CIRCLE  & \CIRCLE &\Circle\\
\citet{yang2023sensefi} & n & \amp & \Circle & \Circle  & \Circle &\Circle\\
\gr \citet{yang2023falldar} & n & BV & \CIRCLE & \CIRCLE  & \Circle &\Circle\\
\citet{zhang2023imgfi}& n & \amp & \Circle & \Circle & \Circle &\Circle\\
\gr \citet{meneghello2023} & ac & \dopp & \CIRCLE & \CIRCLE & \CIRCLE &\Circle\\
\citet{cominelli2023}& ax & \dopp & \CIRCLE & \CIRCLE & \Circle &\Circle\\
\gr \citet{chen2024} & n & \phd & \CIRCLE & \CIRCLE & \Circle &\Circle\\
\citet{hussian2024} & n & \amp & \Circle & \Circle & \Circle &\Circle\\
\midrule
\gr \textit{This work}& ax & \mcr{\amp, \phd\\ \amp\&\phd} & \CIRCLE & \CIRCLE & \CIRCLE & \CIRCLE\\
\bottomrule
\end{tabular}
\begin{tablenotes}[flushleft]
\scriptsize
\item \emph{CSI Data}: Amplitude (\textbf{\amp}), 
Phase Difference (\textbf{\phd}), 
Doppler Shift (\textbf{\dopp}), \item Body Velocity (\textbf{BV}), Short-Time Fourier Transform (\textbf{\stft}). \emph{Cross-Evaluation} \item scenarios: Environments (\faHome), Users (\faUser), Receiver Node Location (\faBroadcastTower),
\item present (\CIRCLE), absent (\Circle). \emph{XAI} Analysis: present (\CIRCLE), absent (\Circle).
\end{tablenotes}
\end{threeparttable}
}
\end{table}
\section{Related Works}
\label{sec:rw}

Research on leveraging \gls{csi} for human sensing has grown rapidly, evolving from coarse metrics to fine-grained subcarrier-level data.
By capturing multipath propagation, \gls{csi} has enabled tasks ranging from presence detection to \gls{har} and vital signs monitoring~\cite{tan2022}.
\gls{dl} has further advanced these capabilities, yet several key challenges remain. 
%
Table~\ref{tab:rw} summarizes representative \gls{har} works by \wifi version, signal representation, cross-evaluation scenarios, and whether explainability was considered.

A key limitation is the lack of native \gls{csi} extraction on commodity devices~\cite{armenta2024wireless}. Early works relied on specialized tools such as the \emph{Linux 802.11n CSI Tool}, thus constraining research to outdated 802.11n hardware.
Although newer frameworks---e.g., \emph{Nexmon} for 802.11ac and \emph{AX-\gls{csi}} for 802.11ax---enable richer data collection, many recent works still rely on legacy tools, thereby limiting their relevance to modern \wifi deployment~\cite{armenta2024wireless}.
Only a few studies leverage the wider bandwidths supported by modern \wifi standards~\cite{meneghello2023,cominelli2023}.
Most \gls{har} approaches rely on \gls{csi} amplitude for its simplicity and real-time suitability~\cite{tan2022}. 
Alternatives such as phase differences, Doppler spectrum, short-time Fourier transform, or body velocity estimates capture richer motion information but increase computational cost, require calibration, and may be affected by orientation and multipath distortions~\cite{tan2022}.
A major challenge across these works is generalization: 
\gls{csi} strongly depends on users, environments, and hardware, often causing severe performance drops when models are tested across different settings~\cite{nirmal2021}.
Although this problem is acknowledged, several studies still evaluate their models in the same context in which they were trained~\cite{yousefi2017,zou2018,yang2022efficient,yang2023sensefi,zhang2023imgfi,hussian2024}. 
In contrast, other works explicitly assess generalization by testing models on users, environments, or device placements that are not seen during training~\cite{wang2019,yang2019,chen2019,zhang2022fewshot,yang2023falldar,meneghello2023,cominelli2023,chen2024}.
To address these challenges, prior works have explored two main strategies. The first involves signal preprocessing to isolate motion-related components and filter out environment-dependent multipath effects~\cite{meneghello2023,yang2023falldar}. While improving stability, these methods may also discard valuable cues together with noise. The second uses learning-based adaptation strategies, e.g., transfer-learning and few-shot learning~\cite{yang2019,chen2024}, which improve cross-domain performance but require additional data collection in each new context, limiting real-world deployability.
Recent evaluations further show that, despite these efforts, generalization across unseen domains remains limited~\cite{cominelli2023}.
Finally, while \gls{dl}-based approaches become the dominant paradigm, they typically behave as black boxes, offering little insight into their decision-making process. 
To the best of our knowledge, no prior works have explored explainability beyond performance metrics. Our work fills this gap by leveraging \sage to systematically analyze model behavior, thus uncovering the intrinsic mechanisms that govern decision-making and paving the way for more transparent 
\wifi sensing systems.

\noindent
\textbf{Positioning.} Our work advances prior research in \emph{three directions}: \ienum{
\item evaluating \gls{har} models under 
standalone and combined \gls{csi} representations (amplitude, phase difference, and their integration via ``early" fusion) using modern \wifi~802.11ax hardware;
\item performing systematic cross-domain evaluations across users, environments, and receiver node locations to assess model generalization; and
\item applying \sage (the core of XAI2CSI) for a comprehensive \gls{xai} analysis, providing a global view of model 
decision mechanisms.}


\section{Methodology}
\label{sec:methodology}

\noindent
\textbf{\gls{csi} background.} 
\gls{csi} characterizes the signal propagation between transmitter and receiver, encapsulating reflection, scattering, diffraction, and multipath effects.
%
It supports communication optimization by enabling tasks such as beamforming and rate adaptation~\cite{tan2022}.
In an \gls{ofdm} system such as \wifi, \gls{csi} provides a detailed estimate of the wireless channel in the frequency domain.
Thanks to the wide bandwidth of \wifi signals and the fine granularity of \gls{ofdm} subcarriers, this information is valuable for both communication optimization and device-free sensing applications, e.g., indoor localization, 
and human activity monitoring.
Formally, in an \gls{ofdm} system with $\nsc$ subcarriers and a single spatial stream, the channel can be modeled in the frequency domain as
$\mathbf{r} = \cfr \otimes \mathbf{s} + \mathbf{n}$,
where $\mathbf{r} \in \complexnum^\nsc$ is the received \gls{ofdm} symbol, $\cfr \in \complexnum^\nsc$ is the channel frequency response (i.e., the \gls{csi}), $\mathbf{s} \in \complexnum^\nsc$ is the transmitted \gls{ofdm} symbol, $\mathbf{n} \in \complexnum^\nsc$ is the noise vector, and $\otimes$ denotes the \emph{Hadamard} product.
%
The channel response $\mathbf{\cfr}$ captures the amplitude attenuation and phase shift of each subcarrier.
The receiver can estimate the channel ($\Cbf \simeq \cfr$) by computing the ratio between the received $\mathbf{r}$ and transmitted $\mathbf{s}$ signals.

In \gls{mimo} systems with $\ntx$ transmit antennas and $\nrx$ receive antennas, the \gls{csi} at time $t$ can be represented as a tensor $\mathcal{\mathbf{H}}(t) \in \mathbb{C}^{\ntx \times \nrx \times \nsc}$, which captures the channel response for every transmit–receive antenna pair across all subcarriers.
Specifically, its amplitude and phase components are defined as: $\amplitude(t)=||\mathcal{\mathbf{H}}(t)||$ and $\phase(t)=\angle{\mathcal{\mathbf{H}}(t)}$, respectively, where $\amplitude(t), \phase (t) \in \mathbb{R}^{\ntx \times \nrx \times \nsc}$.

\gls{csi} is collected by capturing \wifi packets, reflecting both static factors (e.g., room layout) and dynamic ones (e.g., human presence and movement).
While primarily used in transceivers for tasks like equalization and beamforming,
\gls{csi} can also be extracted via dedicated tools. These tools---for each Tx-Rx antenna pair---record the channel estimate across all available subcarriers at a given sampling rate \sr, producing a \gls{csi} time series that reflects wireless channel evolution.


\vspace{2pt}
\noindent
\textbf{\gls{dl}-based \gls{har}.} 
We use \gls{csi} measurements 
across a $1 \times \nrx$ MIMO setup 
to detect human activities, since different actions uniquely perturb wireless propagation~\cite{zou2018}.
We compare \emph{three} input configurations: \ienum{\item amplitude ($\amplitude$) only, \item phase difference ($\phased$) only, and \item their combination ($\amphd$) via \emph{early-fusion}, where both are concatenated as input features.}
To mitigate hardware impairments (e.g., \emph{CFO} or \emph{SFO})~\cite{yousefi2017}, phase differences $\Delta\phi_{j-r} = \phi_j - \phi_r$ are computed across receiving antenna pairs relative to a reference ($R_{x,r}:j \neq r$) and over time via complex conjugate products of raw \gls{csi} samples, preserving relative phase shifts and signal integrity.
%
The resulting time series is segmented into overlapping \emph{frames} $X=\big\{ \bar{\bm{X}}_i \big\}_{i=1}^{N}$, where $\bar{\bm{X}}_i \in \mathbb{R}^{\mathtt{D} \times \nsc \times \wsize}$ captures dynamics over a window $\wsize$.\footnote{With \amphd, the first amplitude sample is discarded to ensure temporal alignment with the $\phased$ series.} 
The spatial dimension $\mathtt{D}$ is $\nrx$ for $\amplitude$, $\nrx$-$1$ for $\phased$, and $2\nrx$-$1$ for $\amphd$. 
Each frame $\bar{\bm{X}}_i$ is fed into a \gls{dl} model \ensuremath{\mathcal{M}_{\bm{\theta}}: \bar{\bm{X}}_i \rightarrow [0,1]^{|\mathcal{C}|}} that outputs activity class probabilities over $\mathcal{C}$.
The predicted label is obtained as \ensuremath{\hat{y} = \arg \max_{c \in \mathcal{C}} \; \mathcal{M}_{\bm{\theta},c}(\bar{\bm{X}}_i)}, where $\mathcal{M}_{\bm{\theta},c}(\bar{\bm{X}}_i)$ is the probability assigned to class $c$.
%
Model training optimizes $\bm{\theta}$ using the \emph{Categorical Cross-Entropy} loss $\mathcal{L}(y,\hat{y})=-\sum_{j=1}^{|\mathcal{C}|}\mathbf{1}_{[y=j]}\ln\big(\mathcal{M}_{\bm{\theta},j}(\bar{\bm{X}}_{i})\big)$,
where $y\in\mathcal{C}$ is the true label.

\vspace{3pt} 
\noindent
\textbf{Interpreting \gls{csi}-based Human Sensing via \sage.} 
%
%
We evaluate the importance of \gls{csi} components using \emph{SAGE}~\cite{covert2020sage}, a model-agnostic approach that, unlike local methods (e.g., LIME, SHAP), quantifies global loss degradation across entire distributions.
\sage quantifies the (global) contribution of each feature by comparing the model’s loss $\mathcal{L}(\cdot,\cdot)$ with and without the feature, thereby measuring its impact on predictive performance.
Formally, for a model $\mathcal{M}(\cdot)$ predicting label $y$ from features $\bm{x}$ (the vectorized frame $\bar{\bm{X}}_i$ in our case), \sage evaluates performance on different feature subsets $\bm{x}_S$ for $S \subseteq \mathcal{D}$, where $\mathcal{D}$ is the set of all feature indices.  
The predictive power $\mathcal{M}(\cdot)$ derives from a subset $\bm{x}_S$ is defined by the function \ensuremath{v_{\mathcal{M}}:\mathcal{P}(\mathcal{D})\rightarrow\mathbb{R}}:
\begin{equation}
v_{\mathcal{M}}(S)=\mathbb{E}[\mathcal{L}(\mathcal{M}_{\varnothing}(\bm{x}_\varnothing),Y)]-\mathbb{E}[\mathcal{L}(\mathcal{M}_{S}(\bm{x}_{S}),Y)]
\end{equation}
%
%
Here, \ensuremath{\mathcal{M}_S(\bm{x}_S) = \mathbb{E}[\mathcal{M}(\bm{X}) \mid \bm{X}_S = \bm{x}_S]} is the conditional expectation denoting the model’s prediction when only features in $S$ are known, marginalizing over the rest.
Conversely, \ensuremath{\mathcal{M}_{\varnothing}(\bm{X}_{\varnothing}) = \mathbb{E}[\mathcal{M}(X)]} is the mean prediction across all features.
%
The \emph{SAGE value} $\phi_i(v_{\mathcal{M}})$ is the Shapley value of the predictive power function $v_{\mathcal{M}}$, attributing a contribution to each feature $x_i$ to the overall model predictive performance. \sage values represent the expected per-instance SHAP values computed w.r.t. the loss $\mathcal{L}$.
%
For a specific sample $(\bm{x}, y)$, this is expressed as 
\ensuremath{v_{\mathcal{M}}^{(\bm{x},y)}(S)=\mathcal{L}(\mathcal{M}_{\varnothing}(\bm{x}_{\varnothing}),y)-\mathcal{L}(\mathcal{M}_{S}(\bm{x}_{S}),y)}
which quantifies how the prediction loss changes when conditioning on the subset of features $S$ compared to using no features at all. 
Finally, the \emph{global importance} of each feature is then obtained by averaging these per-instance contributions across the entire dataset \ensuremath{\phi_{i}(v_{\mathcal{M}})=\mathbb{E}_{\bm{X}Y}[\phi_{i}(v_{\mathcal{M}}^{(\bm{X},Y)})]}.

\section{Experimental Setup}
\label{sec:setup}
This section details the experimental setup: the dataset, pre-processing, the DL model, and the evaluation procedure.

\vspace{2pt}
\noindent
\textbf{Dataset.} 
We use the 
dataset from~\cite{cominelli2023}, collected with the 
\texttt{AX-CSI} tool
supporting the IEEE 802.11ax standard. Each trace records a specific user performing a given activity for $80\unit{s}$ in a particular environment on a given day. 
We restrict our analysis to $5$ activities (Empty Room, Jumping, Running, Sitting, Walking) covering static, repetitive, and dynamic Doppler profiles.
%
The collection was performed over two consecutive days in three furniture-rich environments (\textit{Lab}, \textit{Office}, and \textit{Hall})
involving three users $U \equiv\{U_k\}_{k=1}^{3}$, enabling evaluation across different settings.
\gls{csi} was captured in the $5\unit{GHz}$ band at \ensuremath{\sr = 150\unit{Hz}} via a $1\times 4$ \gls{simo} setup ($\{R_{x,j}\}_{j=1}^4$) across three receiver nodes $\mathcal{N} \equiv \{N_i\}_{i=1}^{3}$ ($\le 160\unit{MHz}$ bandwidth, $\le 2048$ subcarriers/antenna).

\vspace{2pt}
\noindent
\textbf{Pre-processing.}
Following~\cite{cominelli2023}, we apply a series of trace-level processing steps, before feeding the \gls{csi} data into the \gls{dl} model:
\ienum{
\item \gls{csi} data is downsampled to $80\unit{MHz}$ channels with $\nsc=1024$ subcarriers by selecting central frequency bins, exploiting the redundancy in high-resolution \gls{csi} and reducing computational load,
and non-informative subcarriers are discarded using a predefined filter
(\ensuremath{\nsc}: 1024$\to$994).
\item Values are normalized per receiver antenna by their mean amplitude.
\item For amplitude features, a \emph{Hampel filter}~\cite{armenta2024wireless} is applied along the time dimension to mitigate outliers.
\item The 
time series is segmented into overlapping windows of \ensuremath{\wsize} = 50 packets (\ensuremath{\Delta\approx333\unit{ms}}) with a stride of $25$ (i.e., 50\% overlap).}
Finally, Min–Max scaling is applied to the receiver antenna, with parameters computed solely from the training set.

\vspace{2pt}
\noindent
\textbf{Model Considered.} We employ a \texttt{2D-CNN} consisting of three 2D convolutional layers 
with kernel sizes $7$, $(5,3)$, and $(3,3)$ and strides $(3,1)$, $(2,2)$, and $1$, respectively. Each layer 
has ReLU activation and is followed by a $2 \times 2$ max-pooling layer. 
The output of the convolutional block is flattened and passed through a dense layer with $128$ units and ReLU activation, followed by a softmax-activated dense layer for classification with $\mathcal{C}$ units.
To regularize training and reduce overfitting, a dropout (rate of $0.3$)
is applied after both the last pooling and the first dense layer.
%
The frame $\bar{\bm{X}}_i$ is structured as a $(\mathtt{D}, \nsc, \wsize)$ tensor, treating $\mathtt{D}$ as the spatial-feature dimension. This arrangement enables the \texttt{2D-CNN} to capture joint correlations across antennas, subcarriers, and time, while facilitating the interpretability of individual components. 
Following~\cite{yang2023sensefi}, we adopt this architecture for its optimal accuracy-complexity trade-off compared to more complex models (e.g., ResNets, ViT), ensuring real-time execution on 
edge devices.

\vspace{2pt}
\noindent
\textbf{Evaluation Procedure.} To ensure reproducible evaluation while preserving causal relationships, each trace is split along the time axis
into training and test with an $80/20\%$ ratio.
We further reversed $20\%$ of the training set for validation.
%
%
Model training was performed for up to $200$ epochs
using the SGD optimizer.
To compute \sage values, each frame is flattened ($\mathcal{D} = \{1, \dots, \mathtt{D} \times \nsc \times \wsize\}$) and conditional expectations are approximated using a background of $40$ training samples/class with a convergence threshold of $0.05$.
This preserves the original frame structure, enabling the grouping of features by subcarrier, antenna, or time index for interpretable analysis.
We release our code at \url{https://github.com/IdioGuarino/XAI2CSI}.

\begin{figure*}[htb]
\centering
\subfloat[\textnormal{\emph{Cross-user}}\label{fig:bar_cross_users}]{\includegraphics[trim=0 27 0 17, clip, height=.17\textwidth]{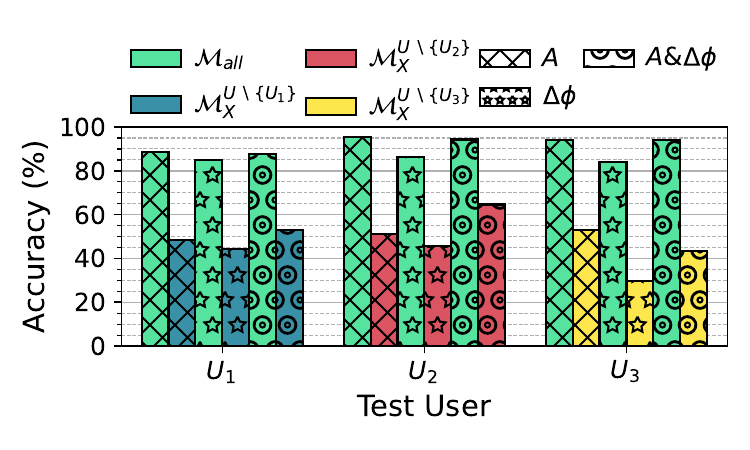}}
\subfloat[\textnormal{\emph{Cross-env}}\label{fig:bar_cross_envs}]{\includegraphics[trim=53 27 0 17, clip, height=.17\textwidth]{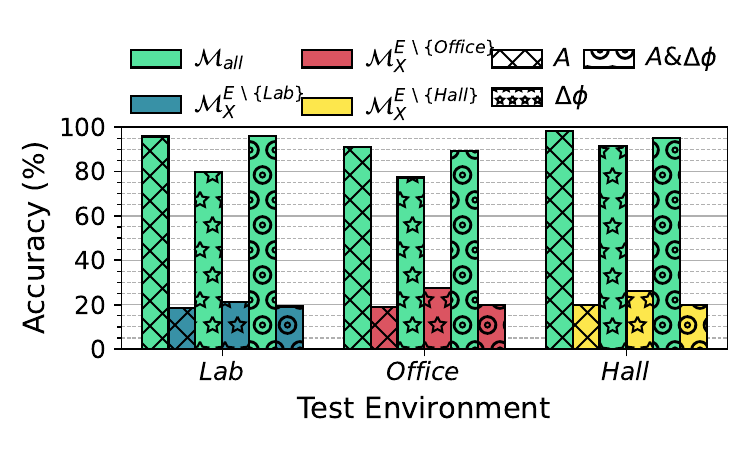}}
\subfloat[\textnormal{\emph{Cross-node}}\label{fig:bar_cross_recvs}]{\includegraphics[trim=53 33 0 17, clip, height=.18\textwidth]{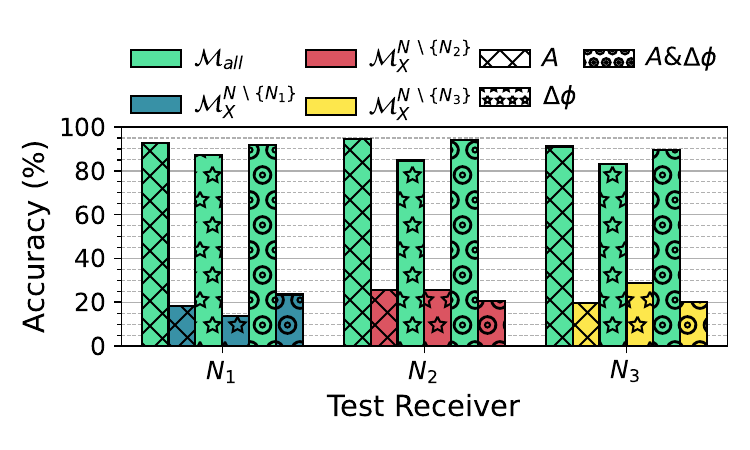}}

\caption{\label{fig:bar_cross} 
$\mcross^{\ast}$ vs. $\mall$ under (a) \textnormal{\emph{cross-user}}, (b) \textnormal{\emph{cross-env}} and, (c) \textnormal{\emph{cross-node}} shifts based on input type: \emph{amplitude} (\amplitude), \emph{phase difference} (\phased), and \emph{their combination} (\amphd). 
Superscript in $\mcross^{\ast}$ denotes training set composition.
$x$\textbf{-axis}: tested user, environment, or node.
}
\end{figure*}
\section{Experimental Evaluation}
\label{sec:evaluation}






In this section, we first evaluate model performance and generalization under contextual changes. Then, we employ \sage to investigate how input features influence decisions.

\begin{figure}[htb]
    \centering

\subfloat[
\label{fig:cm_all_u2_amp}
\centering $A$]{\includegraphics[trim=0 70 0 10, clip, width=.40\columnwidth]{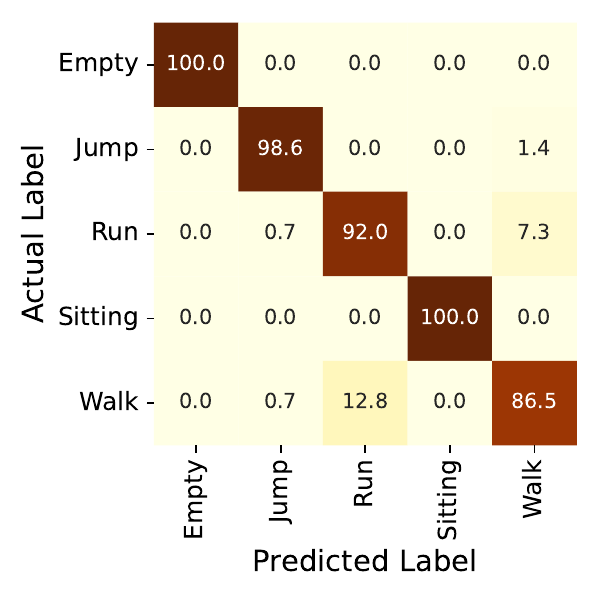}}
\subfloat[
\label{fig:cm_all_u2_ph}
\centering $\Delta \phi$]{\includegraphics[trim=70 70 0 10, clip, width=.30\columnwidth]{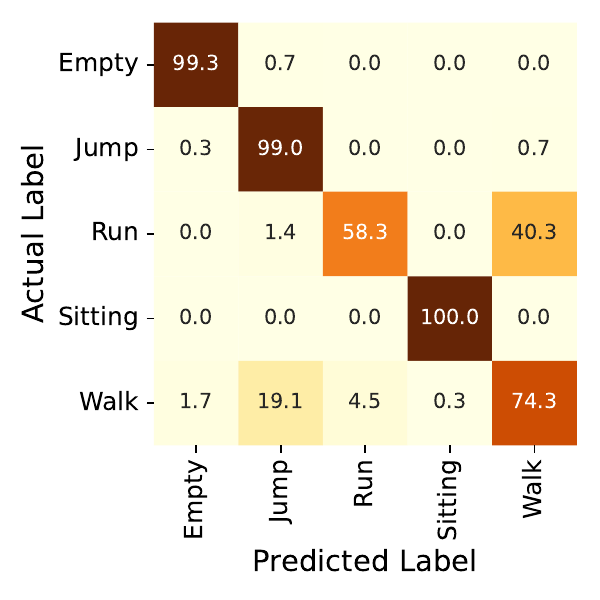}}
\subfloat[
\label{fig:cm_all_u2_amph}
\centering $A \& \Delta\phi$]{\includegraphics[trim=70 70 0 10, clip, width=.30\columnwidth]{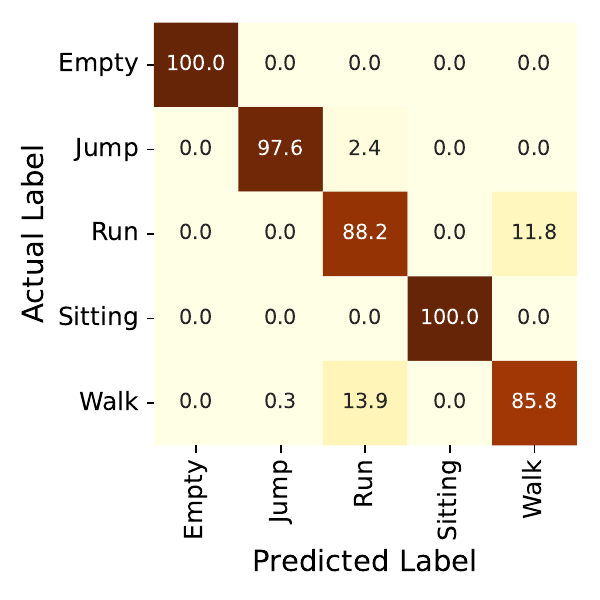}}

\subfloat[
\label{fig:cm_u2_amp}
\centering $A$]{\includegraphics[trim=0 0 0 10, clip, width=.40\columnwidth]{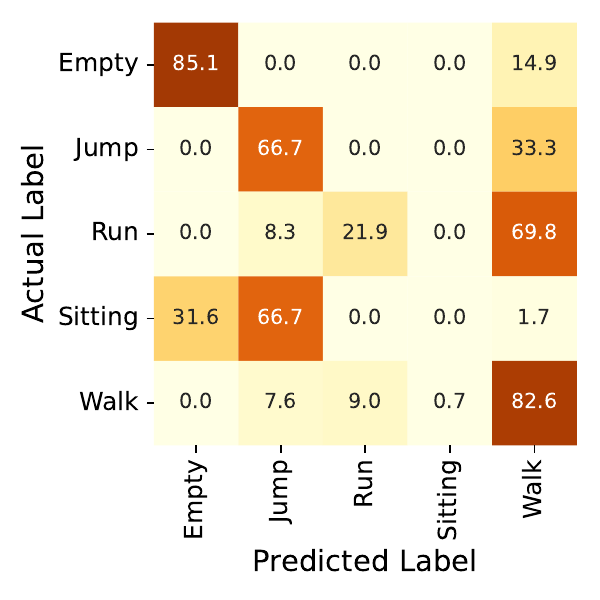}}
\subfloat[
\label{fig:cm_u2_ph}
\centering $\Delta\phi$]{\includegraphics[trim=70 0 0 10, clip, width=.30\columnwidth]{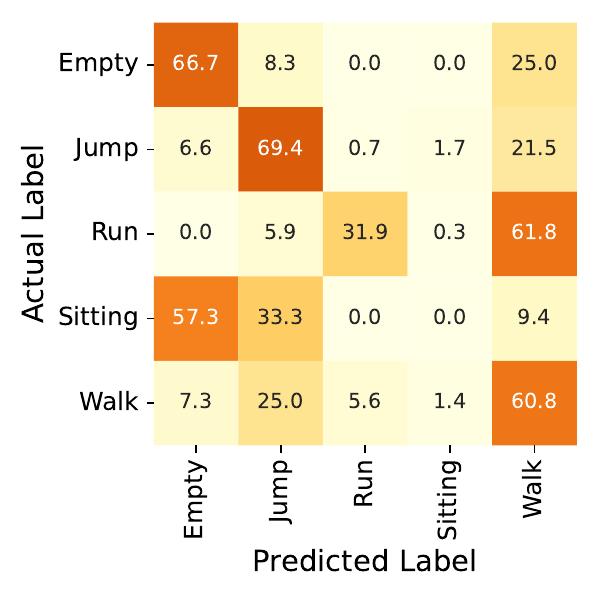}}
\subfloat[
\label{fig:cm_u2_amph}
\centering $A \& \Delta\phi$]{\includegraphics[trim=70 0 0 10, clip, width=.30\columnwidth]{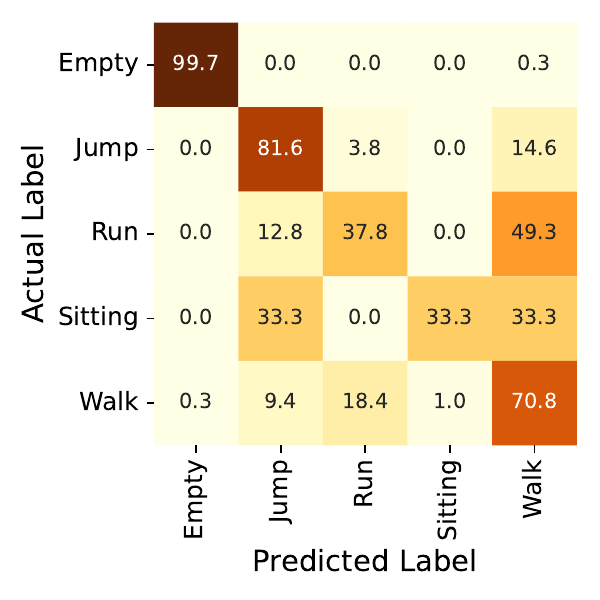}}

\caption{\label{fig:cm_cross}Misclassification errors of $\mall$ (\emph{top}) and $\mcross^{U \setminus \{U_2\}} \triangleright U_2$ (\emph{bottom}) when testing on $U_2$ 
based on input type.
}
    
\end{figure}

\vspace{2pt}
\noindent
\textbf{Assessing Model Performance under Context-Shifts.}
We evaluate our \gls{csi}-based sensing under \emph{three} context shifts: \ienum{\item unseen users (\emph{cross-user}), \item unseen environments (\emph{cross-env}), and \item changes in 
node
position (\emph{cross-node}).} 
In each case, we adopt a leave-one-out cross-validation scheme, where a model trained on all but one setting (viz. $\mcross^{\ast}$) is tested on the unseen one and compared to a model trained on all settings (viz. $\mall$),  serving as an upper bound.

Fig.~\ref{fig:bar_cross} compares the accuracy of $\mcross^{\ast}$ and $\mall$ across the above conditions based on input type: \textit{amplitude} ($A$), \textit{phase difference} ($\Delta\phi$), and \textit{their \emph{early-fusion} combination} ($A\&\Delta\phi$). 
\emph{Cross-user} (resp. \emph{cross-env}) results are obtained by focusing on the \emph{Lab} environment (resp. user $U_1$) while including all nodes. 
In \emph{cross-node}, two nodes are used for training and the remaining one for testing, using data from user $U_1$ in the \textit{Lab}, thus emulating an unseen node location.
As expected, $\mall$ models consistently achieve high accuracy ($88$-$98\%$), with amplitude-based inputs ($A$ and $A\&\Delta\phi$) outperforming the phase-only counterpart ($\Delta\phi$).
Conversely, for $\mcross^{\ast}$ models, performance degradation heavily depends on both the context shift and the input type.
Under \emph{cross-user} conditions, despite the same environment, the accuracy of $\mcross^{\ast}$ drops to at most $65\%$ ($U_2$), with amplitude-based models 
($A$ and $A\&\Delta\phi$) 
yielding the highest robustness.
However, this trend inverts under \emph{cross-env} and \emph{cross-node} conditions, where phase-based models ($\Delta\phi$ and $A\&\Delta\phi$) achieve slightly better performance, despite not exceeding $28\%$ in \emph{cross-env} (\textit{Office}).
This underscores that amplitude is more sensitive to environmental changes and transmitter-receiver geometry, whereas phase difference offers greater cross-domain stability.
%

%
Focusing on \emph{cross-user} settings, Fig.~\ref{fig:cm_cross} shows the confusion matrices for $\mall$ and $\mcross^{U \setminus \{U_2\}}$ models on the unseen user $U_2$, highlighting misclassification patterns across the different input types.
For $\mall$, \emph{Empty} and \emph{Sitting} activities are consistently recognized, while more dynamic activities such as \emph{Run} and \emph{Walk} are often misclassified as each other, especially when leveraging only phase difference (Fig.~\ref{fig:cm_all_u2_ph}).
Conversely, $\mcross^{\ast}$ performance highly depends on the input type.
When considering $A$ or $\Delta\phi$ standalone, \emph{Run} and \emph{Sitting} prove to be the most challenging classes (confused with \emph{Walk} and \emph{Jump}, respectively), with the latter never correctly classified.
Notably, with the same inputs, the model frequently misclassifies \emph{Empty}, which is particularly critical as it implies mistaking an absent user for one performing an activity.
Interestingly, this misclassification pattern is resolved for \emph{Empty}, which the model with $A\&\Delta\phi$ recognizes with near-perfect accuracy. Conversely, for \emph{Run} and \emph{Sitting}, early fusion only partially mitigates the issue, as their classification remains challenging.


\vspace{3pt}
\begin{takehomebox}
\emph{\textbf{Take-Away}: Cross-domain generalizability heavily depends on the input representation. While standalone features suffer severe degradation under context shifts, their early fusion only partially mitigates the issue, resolving macroscopic errors (e.g., empty environments) while leaving dynamic and complex static activities an open challenge.
}
\end{takehomebox}


\vspace{2pt}
\noindent
\textbf{Analyzing feature effects on models' behavior with 
\sage.}
To gain insight into the behavior of the models and understand the reasons behind their failures in cross-user scenarios, we analyze the feature importance for each target user $U_k$, considering two subsets: 
$\bothclass$ containing samples correctly classified by both $\mall$ and $\mcross^{\ast}$, and $\onlyclass$ containing samples correctly classified by $\mall$ but misclassified by $\mcross^{\ast}$. 
This separation allows us to distinguish features that consistently support correct predictions from those associated with failures of $\mcross^{\ast}$. 
Then, we analyze how the different dimensions of the input frame, i.e., receiving antennas, packet arrival order, and subcarriers, affect the models’ decisions.
\begin{figure}[tb]
\centering

\subfloat
{\includegraphics[trim=0 28 0 21, clip, width=.85\columnwidth]{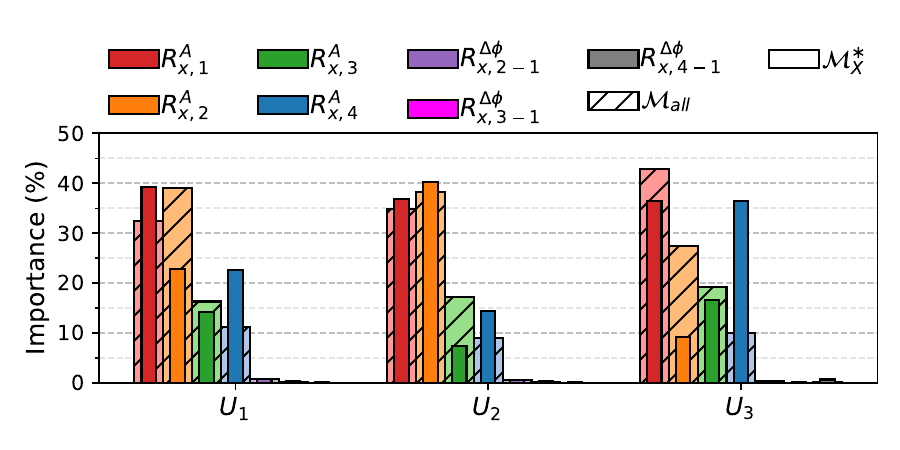}}

\subfloat
{\includegraphics[trim=7 14 0 60, clip, width=.855\columnwidth]{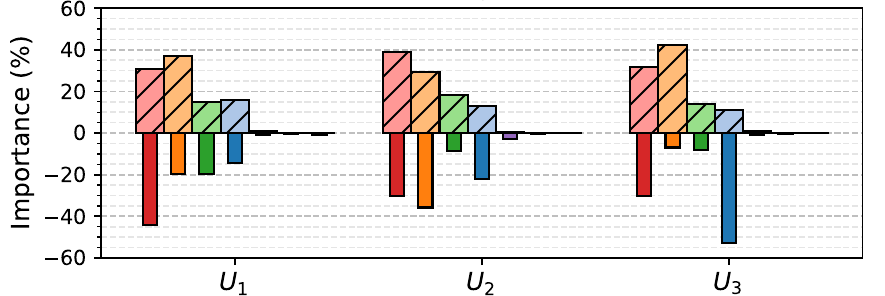}}

\caption{\label{fig:sage_rxs} 
Importance of receiving antenna $R_{x,j}$ ($A\&\Delta\phi$ input) for $\mall$ and $\mcross^{U \setminus \{U_k\}}$ on $\bothclass$ (\textbf{top}) and $\onlyclass$ (\textbf{bottom}) samples across target users $U_k$. 
$\mathbf{R}_{x, j}^A$: \emph{amplitude}. $\mathbf{R}_{x, j-1}^{\Delta \phi}$: \emph{phase difference} w.r.t $R_{x,1}$.
}
    
\end{figure}




    

Fig.~\ref{fig:sage_rxs} compares the relative importance
of the receiving antennas $\{R_{x,j}\}_{j=1}^4$ (normalized by the sum of their absolute values) for the combined-input ($A\&\Delta\phi$) variants of $\mall$ and $\mcross^{U \setminus \{U_k\}}$. Results are broken down by the target user $U_k$, isolating the individual contributions of amplitude 
$\{R_{x,j}^A\}_{j=1}^4$
and phase difference 
$\{R_{x,j-1}^{\Delta\phi}\}_{j=2}^4$.

Specifically, $\mall$ (viz., hatched bars) exhibits a consistent behavior across users and sample sets.
Specifically, the model maintains a stable feature hierarchy in which amplitude components are most important, while phase differences are negligible, with values remaining close to zero. Among the amplitudes, $R_{x,1}^A$ and $R_{x,2}^A$ stand out as the most important, while $R_{x,4}^A$ consistently emerges as the least influential.
This overall trend reveals that the model exploits amplitudes as a leaky shortcut, leading to cross-domain overfitting. In fact, while amplitude provides more stable patterns than volatile phase difference, it remains strictly dependent on the specific user and environment.
Consequently, for $\mcross$, the antennas show a highly unstable behavior depending on the unseen user. For instance, for $U_1$, $\mcross$ over-relies on $R_{x,1}^A$, driving the misclassifications on $\onlyclass$, while underweighting $R_{x,2}^A$, i.e., the most relevant feature for $\mall$. 
Similarly, for $U_3$, while $R_{x,1}^A$ and $R_{x,4}^A$ support correct predictions in $\mcross$, $R_{x,4}^A$ mainly drives its misclassifications, despite consistently representing the least influential structural component for $\mall$.

\begin{figure}[tb]
\centering

\subfloat
{\includegraphics[trim=0 44 0 0, clip, width=.8\columnwidth]{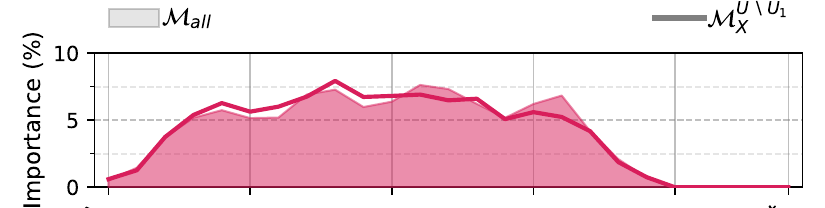}}

\vspace{-5pt}
\subfloat
{\includegraphics[trim=9 5 0 16, clip, width=.8\columnwidth]{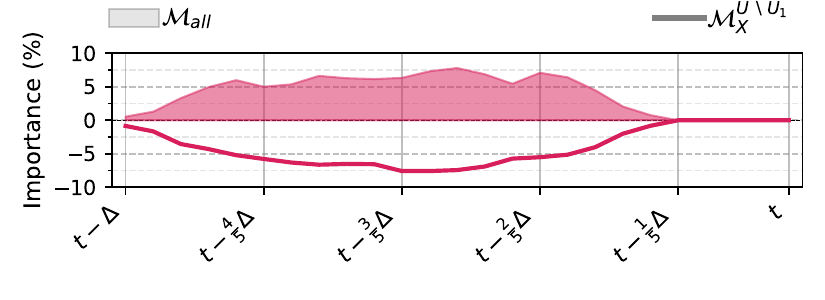}}

\caption{\label{fig:sage_time} Packet arrival order importance ($A\&\Delta\phi$ input) for $\mall$ and $\mcross^{U \setminus \{U_1\}}$ on $\bothclass$ (\textbf{top}) and $\onlyclass$ (\textbf{bottom}) samples of user $U_1$. $x$-axis: chronologically sorted packet indices (2 packets/bin; $t$ is most recent). $\mall$ (areas) vs. $\mcross^{U \setminus \{U_k\}}$ (lines).}
    
\end{figure}

Focusing on $U_1$ for brevity, as other users yield similar results,
Fig.~\ref{fig:sage_time} shows feature importance along the time-axis for $\mall$ and $\mcross^{U \setminus \{U_1\}}$ on $\bothclass$ and $\onlyclass$, computed over consecutive packet pairs within a frame duration $\wsize$. 
In $\bothclass$, both models mainly rely on earlier (i.e., left-handed) or central packets, indicating that \gls{har} is driven by stable motion patterns rather than recent observations, which do not aid accurate predictions.
Conversely, in $\onlyclass$, packets that previously contributed positively now negatively impact $\mcross^{U \setminus \{U_1\}}$.
This indicates that $\mcross^{U \setminus \{U_1\}}$ misinterprets temporal patterns that $\mall$ correctly utilizes, showing its difficulty in generalizing to unseen users.

\begin{figure}[tb]
\centering

\subfloat
[\label{fig:sage_users_subcarriers_both}\centering $\bothclass$]
{\includegraphics[trim=0 30 0 5, clip, width=.9\columnwidth]{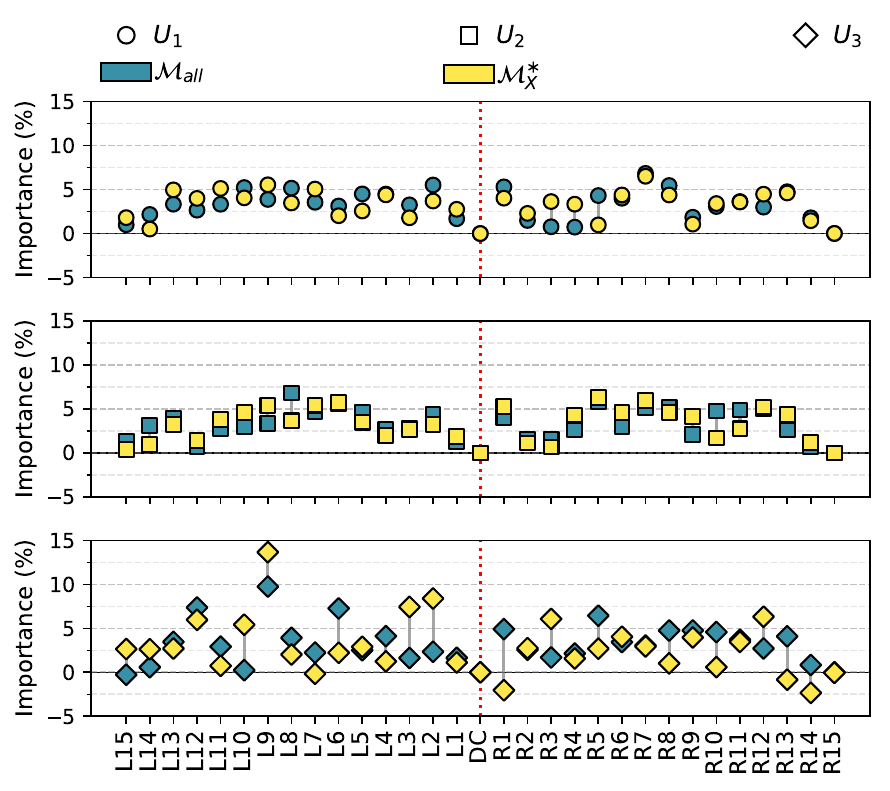}}

\vspace{-9pt}
\subfloat[
\label{fig:sage_users_subcarriers_only}
\centering $\onlyclass$]{\includegraphics[trim=3 5 0 43, clip, width=.915\columnwidth]{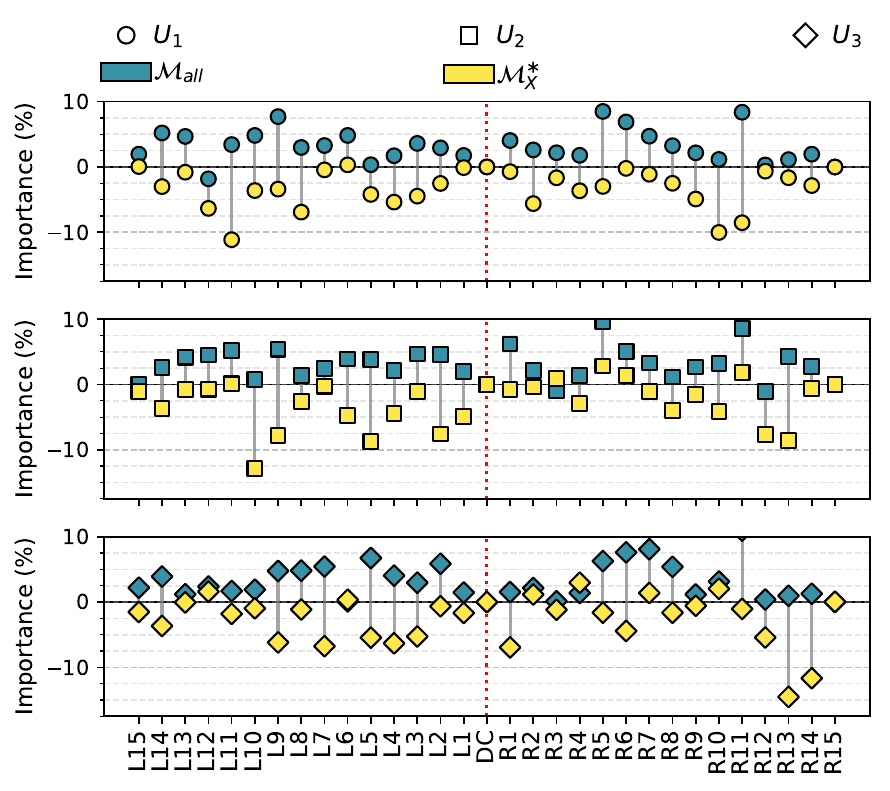}}

\caption{\label{fig:sage_subcarriers} Subcarrier importance ($A\&\Delta\phi$ input) for $\mall$ and $\mcross^{U \setminus \{U_k\}}$ on (a) $\bothclass$ and (b) $\onlyclass$ samples across target users $U_k$ (marked differently).
Left/right subcarriers are grouped into $15$ bins ($\{\mathtt{L}_s\}_{s=1}^{15}$ and $\{\mathtt{R}_s\}_{s=1}^{15}$) around the DC carrier (\emph{dotted red line}).
}
    
\end{figure}

Lastly, Fig.~\ref{fig:sage_subcarriers} shows 
subcarrier importance for $\mall$ and $\mcross^{U \setminus \{U_k\}}$ on $\bothclass$ and $\onlyclass$, computed by grouping left and right subcarriers of the central DC carrier into $15$ bins each.
Specifically, on $\bothclass$, importance values are broadly distributed across bins with most subcarriers contributing positively in both models and ranging in $0$-$15\%$. This indicates that successful classifications rely on information extracted from a wide spectral range.
Moreover, for $U_3$, importance is more dispersed across subcarriers, revealing greater disagreement between models even on correctly classified samples.
In contrast, the analysis on $\onlyclass$ shows that, while $\mall$ maintains positive, distributed importance, $\mcross^{\ast}$ assigns predominantly negative importance to nearly all subcarriers. This indicates that $\mcross^{\ast}$ misinterprets frequency-domain information, perceiving spectral features as misleading and thus driving misclassifications.
Across all users and sample sets, the DC subcarrier has no effect on the decisions of both models.

\vspace{3pt}
\begin{takehomebox}
\emph{\textbf{Take-Away}: 
The analysis indicates that $\mcross^{\ast}$ struggles to generalize to unseen users by over-relying on the most domain-dependent amplitudes while ignoring the potentially more generalizable, yet volatile, phase differences. Conversely, $\mall$ effectively utilizes information across antennas, packets, and subcarriers, whereas $\mcross^{\ast}$ misinterprets these structural signals, leading to severe misclassifications.
}
\end{takehomebox}


\section{Discussion and Conclusions}
\label{sec:conclusions}

This paper presented XAI2CSI, 
a framework embedding \sage as its core to systematically apply \gls{xai} to \gls{csi}-based \gls{har} and investigate the behavior of \gls{dl} models under realistic conditions.
%
Using different \gls{csi} input representations (amplitude, phase difference, and their combination) collected from modern IEEE 802.11ax hardware, we first evaluated model generalization to unseen users, environments, and receiver positions, showing that performance drops significantly when the deployment context changes. This highlights that cross-domain robustness in \wifi sensing remains an open challenge.
Focusing on the \emph{cross-user} scenario, we 
applied \sage 
to analyze the contribution of temporal, spectral, and spatial \gls{csi} components to model decisions.
Our analysis revealed that generalization failures stem from over-reliance on user-specific patterns, which are misinterpreted when the model encounters a new user.
Overall, this study showed how explainability can provide a principled understanding of \gls{dl} behavior in \wifi sensing, paving the way toward transparent, interpretable, and robust \gls{csi}-based systems.
\emph{Future works} will include diverse \gls{xai} methods and \gls{dl} models, mitigating \gls{csi} artifacts via domain adaptation, and enhancing multimodal fusion for cross-domain robustness.


\bibliographystyle{IEEEtranN}
\begingroup
\footnotesize
\bibliography{IEEEabrv, biblio}
\endgroup

\end{document}